\documentclass[3p,times,12pt]{elsarticle}
\usepackage{amsmath}
\usepackage{braket}
\usepackage{amssymb}
\usepackage{amsfonts}
\usepackage{latexsym}
\usepackage{graphicx}
\usepackage{latexsym}
\usepackage{color}
\usepackage{soul}
\def\beq{\begin{eqnarray}}
\def\eeq{\end{eqnarray}}
\def\ln{\,\mbox{ln}\,}

\def\det{\,\mbox{det}\,}

\def\al{\alpha}
\def\be{\beta}

\def\de{\delta}

\def\la{\lambda}
\def\na{\nabla}
\def\pa{\partial}

\def\ph{\varphi}

\def\Ga{\Gamma}
\def\De{\Delta}

\def\Om{\Omega}

\usepackage{indentfirst} %Parï¿œgrafo

\newcommand{\eq}[1]{(\ref{#1})}

\newcommand{\n}[1]{\label{#1}}
\begin{document}

\title{
Stochastic quantization of a self-interacting nonminimal scalar
field in semiclassical gravity}

\author[add1]{Eduardo Antonio dos Reis}
\ead{eareis@fisica.ufjf.br}
\author[add2]{Gast\~ao Krein}
\ead{gastao.krein@unesp.br}
\author[add3]{Tib\'{e}rio de Paula Netto}
\ead{tiberio@sustech.edu.cn}
\author[add1,add4,add5]{Ilya L. Shapiro}
\ead{shapiro@fisica.ufjf.br}
\address[add1]
{Departamento de F\'{\i}sica, \ ICE, \ Universidade Federal de Juiz de Fora,
36036-330 Juiz de Fora, \ MG, \ Brazil}
\address[add2]{Instituto de F\'{\i}sica Te\'orica, Universidade Estadual
Paulista \\
Rua Dr. Bento Teobaldo Ferraz, 271 - Bloco II, 01140-070,
S\~ao Paulo, SP, Brazil}
\address[add3]{
Department of Physics, Southern University of Science and Technology,
Shenzhen 518055, China}
%% E-mail: tiberio@sustech.edu.cn

\address[add4]{Department of Theoretical Physics, Tomsk State Pedagogical
University,  634061 Tomsk, Russia}
\address[add5]{
National Research Tomsk State University, 634050 Tomsk, Russia}

\begin{quotation}
\begin{abstract}
We employ stochastic quantization for a self-interacting nonminimal
massive scalar field in curved spacetime. The covariant background
field method and local momentum space representation are used to
obtain the Euclidean correlation function and evaluate multi-loop
quantum corrections through simultaneous expansions in the curvature
tensor and its covariant derivatives and in the noise fields. The
stochastic correlation function for a quartic self-interaction
reproduces the well-known one-loop result by Bunch and Parker and
is used to construct the effective potential in curved spacetime
in an arbitrary dimension $D$ up to the first order in curvature.
Furthermore, we present a sample of numerical simulations for $D=3$
in the first order in curvature. We consider the model with
spontaneous symmetry breaking and obtain fully nonperturbative
solutions for the vacuum expectation value of the scalar field and
compare them with with one- and two-loop solutions.
\end{abstract}
\end{quotation}
\vskip 4mm

%%%%%%%%%%%%%%%%%%%%%%%%%%

%%%%%%%%%%%%%%%%%%%%%%%%%%%%%%
%%%%%%%%%%%%%%%%%%%%%%%%%%%%%%
\maketitle

%%%%%%%%%%%%%%%%%%%%%%%%%%%%%%
%%%%%%%%%%%%%%%%%%%%%%%%%%%%%%
\section{Introduction}
\label{int}

The standard perturbative approach to the semiclassical gravity
proved to be fruitful for many applications, e.g., to evaluate the
creation of particles in the primordial
universe~\cite{Abbott:1982hn,Dolgov:1982th,Albrecht:1982mp,Dolgov:1989us}
(see \cite{Mukhanov:2005sc} for an introduction) and to explore the
back-reaction of quantum matter on the vacuum \cite{Fischetti:1979ue},
leading to the historically first Starobinsky model of
inflation~\cite{Starobinsky:1980te, Starobinsky:1981vz}. However,
there are situations in which it is necessary to go beyond the framework
of the perturbative approach. This concerns the quantum field theory
(QFT) based description of Starobinsky inflation
\cite{Pelinson:2003gn,Netto:2015cba} and an alternative Higgs
inflation model~\cite{Bezrukov:2007ep,Bezrukov:2008ej,
Bezrukov:2010jz,Bezrukov:2009db,Barvinsky:2008ia,Barvinsky:2009fy},
where two-loop corrections to the effective potential were shown to
modify the bounds for the Higgs mass. Other situations of similar sort
occur in the problem of high energy (UV) Higgs stability, as discussed
recently in \cite{EliasMiro:2012ay} (see further references therein),
and especially the possibility of the low-energy (IR) running of the
observable cosmological
constant~\cite{Shapiro:2009dh}, which can not be proved or disproved
due to technical limitations of the available theoretical methods.
In this last situation, one has to rely on qualitative considerations
based on covariance and dimensional
arguments~\cite{Babic:2001vv,Shapiro:2000dz,Shapiro:2004ch,Shapiro:2008sf},
which lead to a universal form for the IR running of the cosmological and Newton
constants, such that what remains is a single parameter
to be determined from cosmological~\cite{Fabris1,Fabris2} or
astrophysical~\cite{Rodrigues:2009vf,Toniato:2017wmk,Farina:2011me}
data. This is clearly an unsatisfactory situation, since the possibility
of the IR running of the cosmological constant is not supported by
a solid quantum field theory calculation.

A situation somehow similar to the ones described above takes place in
quantum chromodynamics (QCD), where the perturbative approach is
inefficient in the IR sector of the theory. It is well known that the
framework of
lattice QCD has had remarkable success in e.g. reproducing the masses
of the ground-state light hadrons~\cite{Durr:2008zz} and has
become the standard source of insight into the IR sector of QCD.
In the lattice approach the theory is reformulated on a discrete Euclidean spacetime lattice
and solved as a classical statistical mechanics problem employing
Monte-Carlo methods~\cite{Gattringer:2010zz}. Furthermore, when
Monte-Carlo methods become inefficient, a most-used method is
stochastic quantization, which also provides a promising alternative
to deal with complex-action problems~\cite{Aarts:2015tyj}. In stochastic
quantization, quantum fluctuations are obtained with the use of the Langevin
equation. A first attempt to derive the Schr\"odinger equation within such an
approach is due to Nelson~\cite{Nelson:1966sp}, which has proven to be
an inadequate quantization procedure due to the use of a classical noise source.
Its modern incarnation dates back to the 1980's with the work of Parisi
and Wu~\cite{Parisi:1980ys}, who found the correct quantization procedure,
in that quantum mechanics is obtained as the thermal equilibrium limit of a
hypothetical stochastic process with respect to a fictitious time (or quantum
dimension) variable~$\tau$ for the evolution of the Langevin equation.
Within this approach, an Euclidean quantum field theory in $D$ dimensions is
obtained as the equilibrium limit for $\tau \rightarrow \infty$ of a classical
system in $D+1$ dimensions coupled to a random noise source with
strength~$\hbar$. When applied to gauge theories, stochastic
quantization is useful to resolve certain issues related to gauge
fixing ambiguities. Refs.~\cite{Damgaard:1988nq,Namiki:1993fd} are
reviews for early developments and applications of this approach.

Given its nonperturbative, Lorentz invariant nature, stochastic
quantization is an alternative to the functional integral and it
seems natural to formulate the approach in an arbitrary curved
spacetime, dealing with the semiclassical approach to gravity.
We note that there is a vast literature considering stochastic
processes in semiclassical gravity, the most well known being
the theory of stochastic gravity (SG); a thorough review with an
extensive list of references on SG can be found in
Ref.~\cite{Hu:2008rga}. Stochastic processes enter in SG by treating
the back reaction of the quantum fields on the spacetime geometry
within the framework of an open quantum system~\cite{Hu:1989db},
in that the quantum fields act as the environment and the gravitational
background as the system. In this case, the stochastic process
is a physical one. One can also mention the important work of
Ref.~\cite{StarYok}, where the correlation functions for a massive
nonminimal scalar field were considered on de Sitter background
by means of the Fokker-Planck equation. In stochastic quantization,
on the other hand, the stochastic process is driven by a fictitious
time variable and provides a third method of
quantization~\cite{Namiki:1993fd}, different from the canonical and
path-integral methods, but tailored for nonperturbative calculations
through numerical simulations. More recently there have been
suggestions~\cite{Polyakov:2000xc,Mansi:2009mz,Oh:2012bx}
that it might be closely related to the holographic principle, as both
state an equivalence between a $D$-dimensional quantum theory and
$(D + 1)-$dimensional classical theory.

In the present work we begin the systematic consideration of
stochastic quantization of semiclassical gravity in an arbitrary
curved spacetime. Although stochastic quantization can be employed
to consider the back reaction problem discussed
above~\cite{Damgaard:1988nq}, the focus of interest here is limited
to the quantum aspects of matter fields propagating in a given
background curved  spacetime.  Starting from the covariant form of
the Langevin equation for a massive self-interacting nonminimal
scalar field in curved spacetime, we develop a computational scheme
based on a local momentum representation to obtain multi-loop quantum
corrections to the classical solution. The multi-loop corrections are
obtained through a simultaneous expansion in the curvature tensor~$R$
and in $\hbar$. In this respect, our approach is different from
earlier~\cite{Damgaard:1988nq}
and more recent~\cite{Menezes:2007rr,deAguiar:2008km} applications of
stochastic quantization to quantum gravity, in that the use of the local
momentum representation allows us to use stochastic quantization as in
flat spacetime. To test the formalism analytically, we first derive the
autocorrelation function of the scalar field and the effective potential
to first order in the curvature tensor~$R$ and in $\hbar$. The results for
the correlation function and effective potential are exactly the same as
derived by other methods, confirming the correctness of the covariant
formulation of the Langevin equation. Next, we consider the vacuum expectation
value of the scalar field for the case of spontaneous symmetry breaking.
We solve analytically and numerically (for $D=3$) the order$-R$ Langevin
equation at ${\cal O}(\hbar)$ (one-loop) and ${\cal O}(\hbar^2)$ (two-loop).
We also solve numerically the full equation, the one that keeps
the full $\hbar$ dependence, and compare with the solutions at one- and
two-loops.

The paper is organized as follows.
In Sec.~\ref{sect2}
the lowest-order derivation of the correlation function of the
scalar field theory nonminimally coupled to gravity is presented.
The calculation is performed analytically by perturbatively
solving the Langevin equation using the Riemann normal
coordinates.  The result is used to obtain the effective potential
to first order in curvature with the cutoff regularization scheme.
In order to test the method and also to generalize a previous result
obtained by means of the local momentum
representation~\cite{Sobreira:2011ep}, present calculation is
performed in an arbitrary Euclidean dimension $D$.
Section~\ref{sect4} presents results of the numerical simulations
in $D=3$. Finally, our Conclusions and Perspectives are
presented in Sec.~\ref{sect5}.

Our notations mainly follow those of Bunch and Parker~\cite{BP}.
The Riemann curvature tensor is defined by
$\,R^{\al} _{\,.\,\be\mu\nu} =
\pa_\mu \Ga^\al_{\be \nu} + \dots\,$
and the Ricci tensor is $R_{\mu\nu} = R^{\al}_{\,.\,\mu\al\nu}$.
For performing quantum calculations, we assume analytic continuation
to Euclidean spacetime, but use the notation $\,\eta_{\mu\nu}\,$
for the flat spacetime metric. The determinant of the metric is
$\,g = \det (g_{\mu\nu})$.

%%%%%%%%%%%%%%%%%%%%%%%%%%%%
%%%%%%%%%%%%%%%%%%%%%%%%%%%%
\section{Correlation
function and effective potential in stochastic quantization}
\label{sect2}

We are interested in
the theory of a nonminimally coupled with gravity self-interacting
massive scalar field $\ph$, with the action
\beq
S = \int  d^Dx\, \sqrt{g} \, \left[
\frac{1}{2}\, g^{\mu\nu} \pa_{\mu} \ph \, \pa_{\nu} \ph
+ \frac{1}{2}\left(m^2 + \xi R \right) \ph^2
+ V(\ph) \right] ,
\label{ep2}
\eeq
where $D$ is the spacetime dimension, $R$ is the scalar curvature
and $V(\ph)$ is a potential, and
$m^2 > 0$ or $m^2 < 0$, the latter being the case of spontaneous
symmetry breaking. $\xi R\ph^2$ is the nonminimal term and $\xi$
is the nonminimal parameter, while $m^2 \ph^2 + V(\ph)$ is the
minimal part of the classical potential. The addition of the
nonminimal term is necessary for the renormalizability of the quantum
theory in $D=4$ (see e.g. Ref.~\cite{book} for
an introduction to the subject).
Although for the formal developments the form of the potential
is not really essential, for the explicit calculations we use the
function $\,V(\ph)={\textstyle \frac{\la}{4}}\,\ph^4$, with $\,\la > 0$.
In the numerical simulations we consider the case of spontaneous
symmetry breaking.

To implement the stochastic quantization, an extra coordinate $\tau$,
the fictitious time, also known as Markov parameter, is introduced
and the scalar field is supplemented with this additional coordinate,
$\ph (x) \rightarrow \ph(x,\tau)$. The dynamics of the field $\ph(x,\tau)$
is described  by a Langevin equation driven by a random white noise
field $\eta(x,\tau)$. Let us start by writing the Langevin equation in
curved spacetime in a covariant form,
\beq
\label{le}
\frac{\pa \ph(x,\tau)}{\pa \tau} =
-\frac{1}{\sqrt{g}}\,
\frac{\de S}{\de \ph(x,\tau)}
\,+\, \eta_c (x,\tau)
\,,
\eeq
where $S$ is the action defined previously in Eq.~\eqref{ep2}
and $\eta_c(x,\tau)$ is the ``covariant" white noise field. By
covariant we mean that the gaussian white noise  obeys the following
correlations (generalized Einstein's relations)
\beq
\langle \eta(x,\tau)\rangle_\eta = 0, \hspace{0.5cm}
\langle \eta(x,\tau)\,\eta(x',\tau')\rangle_\eta
= 2\hbar \,\de_c(x,x') \, \de(\tau - \tau') ,
\label{white-noise}
\eeq
where $\langle{\cdots\rangle}_\eta $ means the stochastic average
and $\de_c (x,x')$ is the covariant delta function, which is
symmetric in the arguments $x$ and $x'$ and also satisfies
\beq
\int d^D y \, \sqrt{g(y)} \, f(y) \, \de_c (x,y) = f(x)\,.
\n{r2}
\eeq
It is not difficult to verify that the solution for this delta function is
$ \de_c (x,x') = g^{-1/4} \, \de^D (x - x') \,g'^{\,-1/4}$,
where \ $g=g(x)$, \ $g'=g(x')$ \  and \ $\de^D (x - x')$ \ is the
ordinary Dirac delta function in flat space.  Although we have been using
and will continue to use $\hbar = 1$ in the following, we have
written it explicity in Eq.~\eq{white-noise} to recall the quantum nature of
the noise fields. The main idea of the Parisi and Wu stochastic
quantization~\cite{Parisi:1980ys} is that quantum field theory vacuum
expectation values $\langle \ph(x_1) \cdots \ph(x_n)\rangle$,
which in the path integral approach are given by
\beq
\langle \ph(x_1) \cdots \ph(x_n)\rangle = \frac{1}{Z}\int {\cal D}\ph \;  \ph(x_1)
\cdots \ph(x_n) \; e^{-S[\ph]},
\label{corr-pi}
\eeq
are obtained from the stochastic averages of correlation functions
of   the field $\ph(x,\tau)$ in the $\tau \to \infty$ limit (the
equilibrium limit). This is because the equilibrium (i.e.
$\tau \rightarrow \infty$ limit) probability distribution of field
configurations generated by the Langevin equation is precisely the
Boltzmann weight $\exp (-S[\ph])$ that enters in Eq.~(\ref{corr-pi}).
This can be proven very easily~\cite{Damgaard:1988nq,Namiki:1993fd}
by considering the Fokker-Planck equation for the probability
distribution associated with the Langevin equation in Eq.~(\ref{le}).
It is important to note that this nonperturbative proof of
equivalence between the quantum Euclidean path integral measure
and the stationary distribution of a stochastic process does not rely
on the white noise nature of $\eta(x,\tau)$; the use of colored noise
fields in a generalized Langevin equation with a positive memory
kernel satisfying the fluctuation-dissipation theorem leads to the
same equivalence~\cite{Damgaard:1988nq}. The use of white noise
is a matter of simplicity, although colored noise naturally provides
an ultraviolet cutoff that regularizes the ultraviolet behavior of the field
theory~\cite{Bern:1985dk}, a welcome feature for the renormalization
program in lattice numerical simulations~\cite{Pawlowski:2017rhn}.
Let us stress that the use of the white noise in a general curved space
is a complicated issue, and here we solve it by using normal
coordinates, that means we define the white noise effectively in
the flat space.

We are interested in evaluating the correlation function $G_c(x,x')$.
This correlation function is obtained as
\beq
\label{fp}
G_c(x,x') =
\lim_{\tau \to \infty} \De(x,x'|\tau) ,
\eeq
where
\beq
\label{ccf}
\De(x,x'|\tau) =
\langle \ph(x,\tau) \, \ph(x',\tau)\rangle_\eta
- \langle \ph(x,\tau)\rangle_\eta
\, \langle \ph(x',\tau)\rangle_\eta.
\eeq
For the action in Eq.~\eq{ep2}, the Langevin equation
Eq.~\eq{le} reads
\beq
\label{le1}
\frac{\pa \ph(x,\tau)}{\pa \tau} =
- \left[-\square + m^2 + \xi R\right]\ph(x,\tau)
- V' \big(\ph(x,\tau)\big) + \eta_c(x,\tau) %.
\,,
\eeq
where $\square = g^{\mu\nu}\na_\mu \na_\nu $ is the d'Alembert
operator. To compare with available results in the literature, obtained
in the context of the loop expansion, we will solve this equation by an expansion
in powers of $\hbar$. From Eq.~(\ref{white-noise}), one can see that
$\eta = {\cal O}(\hbar^{1/2})$. We write for the field $\ph(x,\tau)$ the expansion
\beq
\label{en}
\ph(x,\tau) =
\phi(x) + \ph^{(1)}(x,\tau) + \ph^{(2)}(x,\tau) + \cdots,
\eeq
where $\phi(x)$ is the ${\cal O}(\hbar^0)$ classical background field
and the $\ph^{(n)}(x,\tau)$ are quantum corrections of order
${\cal O}(\hbar^{n/2})$. Using the expansion in the potential,
one obtains
\beq
\n{pot-e}
V' \big(\ph(x,\tau)\big) &=&
V'(\phi)
+ V''(\phi) \, \ph^{(1)}(x,\tau)
+ V''(\phi) \, \ph^{(2)}(x,\tau)
%\nonumber
%&+&
+ \frac12\, V'''(\phi) \, [\ph^{(1)}(x,\tau)]^2 + \cdots ,
\eeq
where dots stand for next order contributions with $n > 2$.
Replacing Eqs.~\eq{en} and \eq{pot-e} into the Langevin equation
Eq.~\eq{le1} and equating terms of the same order in $\hbar$,
one finds the following set of equations for the $\ph^{(n)}$:
\beq
\frac{\pa \phi(x)}{\pa \tau} &=&
- \left[-\square + m^2 + \xi R \right]\phi(x) - V'(\phi) ,
\label{cs}
\\
\frac{\pa \ph^{(1)}(x,\tau)}{\pa \tau} &=&
- \left[-\square + m^2 + \xi R + V''(\phi)\right]\ph^{(1)}(x,\tau)
+ \eta_c(x,\tau) ,
\label{s1}
\\
\frac{\pa \ph^{(2)}(x,\tau)}{\pa \tau} &=&
- \left[-\square + m^2 + \xi R + V''(\phi)\right] \ph^{(2)}(x,\tau)
- \frac{1}{2}\, V'''(\phi)\,[\ph^{(1)}(x,\tau)]^2,
\label{s2}
\eeq
and similarly for $n > 2$. This set is formed by linear equations
which can be, in principle, solved by iteration. We note that a
noise  expansion procedure was used previously in
cosmology~\cite{Martin:2004ba,Martin:2005ir,
Martin:2005hb,Kunze:2006tu} in a different context, not related
to stochastic quantization. Ref.~\cite{Cassol-Seewald:2012tcq}
is first publication related to noise perturbation in the context of
stochastic quantization.

Although Eqs.~(\ref{cs})-(\ref{s2}) are linear equations, they
can not be solved analytically in general when the classical solution
is not a constant. Therefore, for the analytical calculation of
the propagator and effective potential, we take
$\phi(x) \equiv \phi_{\rm cl} = {\rm const}$. The following initial
conditions are assumed:
\beq
\label{ic0}
\ph(x,0)  = \phi(x) + \ph^{(1)}(x,0) + \ph^{(2)}(x,0)
+ \cdots = \phi_{\rm cl},
\eeq
which imply
\beq
\label{ic1}
\ph^{(1)}(x,0) = \ph^{(2)}(x,0) = \cdots = 0 .
\eeq

%%%%%%%%%%%%%%%%%%%%%%%%%%%%
%%%%%%%%%%%%%%%%%%%%%%%%%%%%
\subsection{Correlation function}

Using the noise expansion of the field as in Eq.~\eq{en} in the correlation
function defined in Eq.~\eq{ccf}, one finds for one-loop contribution
\beq
\label{ccf-1}
\De^{{\rm 1-loop}}(x,x'|\tau) =
\langle \ph^{(1)}(x,\tau)\, \ph^{(1)}(x',\tau)\rangle_\eta .
\eeq
Therefore, to obtain the correlation function up to first order in
$\hbar$, one just needs to solve the first order equation given
in Eq.~\eqref{s1}. In order to evaluate this equation, we rely on
the local momentum representation method based on the
Riemann normal coordinates \cite{BP}, which is a useful
formalism for mass-dependent calculations of local quantities,
such as the effective potential~\cite{Sobreira:2011ep}. The
advantage of these special coordinates is that one can use flat-space
methods of calculations, such as momentum representation,
and apply an expansion in powers of the curvature tensor and its
covariant derivatives at the point $P$ to all relevant quantities
{\textemdash} see
e.g. Eqs.~(\ref{ebox}) and (\ref{eboxg}) below.
The method is very efficient especially for deriving local quantities.

In the normal coordinates formalism, the spacetime metric
$g_{\al\be}(x)$ and all related quantities are expanded near a point
$P(x')$, where the metric is supposed to be flat.
%  {\textemdash} this
% is not strictly necessary, albeit it is a useful restriction.
In the first order in the curvature, we have the
dimension-independent result~\cite{Petrov}:
\beq
g_{\al\be}(x) =\eta_{\al\be}
- \frac{1}{3}\,R_{\al\mu\be\nu}(x') \,y^\mu y^\nu
+ \cdots \,,
\label{em}
\eeq
where $x'$ are the coordinates of the point $P$ and
$\,y^\mu=x^\mu-x'^\mu\,$ are deviations from this point.
%%   Objects of our interest are the autocorrelation function
%%   and the effective potential, which are certainly local and,
%%   therefore, the local momentum representation method should
%%   work in any approach, including stochastic quantization.

Before starting the normal coordinates expansion, we follow Bunch and
Parker~\cite{BP} and introduce  a noncovariant correlation function
$\bar{\De}(x,x'|\tau)$ through the relation
\beq
\De(x,x'|\tau) = g^{-1/4}\,\bar{\De}(x,x'|\tau)\, g'^{\,-1/4}\,.
\label{cfd}
\eeq
In the close analogy with Eq.~\eq{ccf-1}, we also introduce a
new field $\bar{\ph}$ such that
\beq
\label{ccn}
\bar{\De}^{1-loop}(x,x'|\tau)
=
\langle \bar{\ph}^{(1)}(x,\tau) \, \bar{\ph}^{(1)}(x',\tau)\rangle\,,
\eeq
where the relation between the old field variable $\ph$ and the new
one $\bar{\ph}$ is given by $\ph = g^{-1/4}\, \bar{\ph}$. Also, by defining
a new noise field $\bar{\eta}$ through $\eta = g^{-1/4}\,\bar{\eta}$,
the noise correlation function reduces to
\beq
\n{mean-flat}
\langle\bar{\eta}(x,\tau) \, \bar{\eta}(x',\tau') \rangle_\eta =
2\, \de^{D}(x-x') \, \de(\tau - \tau') ,
\eeq
where $\de^{D}(x-x')$ is the flat spacetime delta
function.
The crucial point behind this equation is that its {\it r.h.s.} does not
depend on the metric tensor. Hence, after introducing the normal coordinates
one can use the standard flat spacetime stochastic quantization procedure.
In terms of the new variables, after multiplying both sides of Eq.~\eq{s1} from the left
by $g^{1/4}$, one obtains for the ${\cal O}(\hbar^{1/2})$ field equation
\beq
\label{lee1}
\frac{\pa \bar{\ph}^{(1)}(x,\tau)}{\pa \tau} =
- \left[-\,g^{1/4}\, \square \, g^{-1/4}
+ m^2 + \xi R + V''(\phi) \right] \bar{\ph}^{(1)}(x,\tau)
+ \bar{\eta}(x,\tau)\,.
\eeq

Next, we introduce the expansion in normal coordinates and
solve Eq.~\eq{lee1} %Eq.~\eq{G-nc}
order by order in a curvature expansion around
flat spacetime. Regardless the fact that the covariant derivative
commutes with $g$ in general, this is not true order by order
in the expansion in normal coordinates; see e.g. Eqs.~\eq{ebox}
and \eq{eboxg}, where clearly the terms at first order in curvature
tensors are different. Therefore, the presence of the extra factors
of $g^{\pm 1/4}$ in Eq.~\eq{lee1} %Eq.~\eq{1/4g-1/4}
is an important aspect of %our
the approach.

Equation~\eq{lee1} has an appropriate form for introducing
the Riemann normal coordinates. In particular, we are interested in
the terms that are of first order in the curvature. Using Eq.~(\ref{em}),
the expansions of $R$ and $\square$ up to first order in the curvature
are given by
\beq
R(x) &=& R(x') + \cdots , \label{eR}
\\
\square &=& \pa^2
+ \frac{1}{3}\, R^{\mu\,\nu}_{\,\al\,\be}(x')
\,y^\al y^\be \, \pa_\mu \pa_\nu
-\frac{2}{3}\,R^{\al}_{\be}(x') \, y^\be \, \pa_\al
+ \cdots,
\label{ebox}
\eeq
so that
\beq
g^{1/4} \, \square \, g^{-1/4}= \pa^2
+ \frac16\, R
+ \frac{1}{3}\left[ R^{\mu\,\nu}_{\,\al\,\be} (x')
\,y^\al y^\be \, \pa_\mu \pa_\nu
- R^{\al}_{\be} (x') \, y^\be \, \pa_\al \right] + \cdots\,,
\label{eboxg}
\eeq
where the derivatives are $\pa_\al = \pa/\pa y^\al$,
$\,\pa^2 = \eta^{\mu\nu} \pa_\mu \pa_\nu$ and $\cdots$
stands for the terms of higher orders in the curvature and
its covariant derivatives. Deriving the effective potential, one
can safely consider $\,V'' (\phi) = const\,$ and simply replace
$\,m^2\,$ by $\,\tilde{m}^2 = m^2 + V''(\phi)\,$ in Eq.~\eq{lee1}
and in the following. Using Eqs.~\eq{em} and \eq{eboxg} in
Eq.~\eq{lee1}, we obtain
\beq
\label{lee1f}
\frac{\pa \bar{\ph}^{(1)}(x,\tau)}{\pa \tau} &=&
\,-\, \left(-\pa^2 + \tilde{m}^2\right) \bar{\ph}^{(1)}(x,\tau)
- \left(\xi-\frac{1}{6} \,R\right) \bar{\ph}^{(1)}(x,\tau)
\nonumber \\
&+&
\frac{1}{3}\left[
R^{\mu\,\nu}_{\,\al\,\be} \,y^\al y^\be\, \pa_\mu \pa_\nu
- R^{\al}_{\be} \, y^\be \, \pa_\al\right] \bar{\ph}^{(1)}(x,\tau)
+ \cdots +  \bar{\eta}(x,\tau) \,.
\eeq
This equation is the generalized Langevin equation that
we need to solve to obtain the one-loop contribution to the correlation
function. It contains an infinite expansion in the curvature
$R$ but it can be solved consistently order by order within the
curvature expansion by expanding $R$ and its covariant derivatives
at the point $P$.

Proceeding in the same way for all the $\varphi^{(n)}$, i.e. defining $\ph^{(n)} = g^{-1/4}\,
\bar{\ph}^{(n)}$, the expansion in $R$ is obtained by expanding each of the $\bar{\ph}^{(n)}$
in the form
\beq
\label{ce}
\bar{\ph}^{(n)} (x,\tau) = \bar{\ph}^{(n)}_0(x,\tau)
+ \bar{\ph}^{(n)}_1 (x,\tau)
+ \cdots \,,
\eeq
where the superscript $n$ indicates the power of $\hbar$ and the subscripts $0$ and $1$ indicate
the power of $R$ so that $\bar{\ph}^{(n)}_0$ is of ${\cal O}(R^0)$ and $\bar{\ph}^{(n)}_1$
is of ${\cal O}(R)$. We restrict the expansion to first order in~$R$. By direct use
of this expansion into Eqs.~(\ref{cs})-(\ref{s2}) we obtain the following set of equations:
\beq
\label{eq0}
%O(1)
\frac{\pa \bar{\ph}^{(1)}_0(x,\tau) }{\pa \tau} &=&
- \left(-\pa^2 + \tilde{m}^2\right) \bar{\ph}^{(1)}_0(x,\tau)
+ \bar{\eta}(x,\tau)\,,
\\
%O(2)
\frac{\pa \bar{\ph}^{(2)}_0(x,\tau)}{\pa \tau} &=&
- \left(- \pa^2 + \tilde{m}^2 \right) \bar{\ph}^{(2)}_0(x,\tau)
- \frac{1}{2} V^{(III)} \, \left(\bar{\ph}^{(1)}_0(x,\tau)\right)^2
\,,
\\
\label{eq1}
%O(3)
\frac{\pa \bar{\ph}^{(3)}_0(x,\tau)}{\pa \tau} &=&
- \left(-\pa^2 + \tilde{m}^2 \right) \bar{\ph}^{(3)}_0(x,\tau)
-   V^{(III)} \, \bar{\ph}^{(1)}_0(x,\tau) \bar{\ph}^{(2)}_0(x,\tau)
-  \frac16   V^{(IV)}  \, \left(\bar{\ph}^{(1)}_0(x,\tau)\right)^3
\,,
\\
\n{P4}
%O(4)
\frac{\pa \bar{\ph}^{(4)}_0(x,\tau)}{\pa \tau} &=&
- \left(-\pa^2 + \tilde{m}^2 \right) \bar{\ph}^{(4)}_0(x,\tau)
-   V^{(III)} \, \left[ \bar{\ph}^{(1)}_0(x,\tau) \bar{\ph}^{(3)}_0(x,\tau)
+ \frac12 \left(\bar{\ph}_0^{(2)}(x,\tau) \right)^2 \right]
\nonumber \\
&-& \frac12  V^{(IV)} \, \left(\bar{\ph}^{(1)}_0(x,\tau)\right)^2 \bar{\ph}^{(2)}_0(x,\tau),
\eeq
for flat spacetime, and
\beq
\n{O(1)}
\frac{\pa \bar{\ph}^{(1)}_1}{\pa \tau}(x,\tau) &=&
-\, \left(-\pa^2 + \tilde{m}^2\right)\, \bar{\ph}^{(1)}_1(x,\tau)
- \left(\xi-\frac{1}{6}\right) R \, \bar{\ph}^{(1)}_0(x,\tau),
\\
\n{O(2)}
\frac{\pa \bar{\ph}^{(2)}_1(x,\tau)}{\pa \tau} &=&
- \left(- \pa^2 + \tilde{m}^2 \right) \bar{\ph}^{(2)}_1(x,\tau)
- \left(\xi - \frac16 \right) R \, \bar{\ph}^{(2)}_0(x,\tau)
- V^{(III)} \, \bar{\ph}^{(1)}_0(x,\tau) \bar{\ph}^{(1)}_1(x,\tau) \nonumber \\
&-& \frac{1}{24}  V^{(III)} \, R_{\mu\nu} y^\mu y^\nu\, \left(\bar{\ph}^{(1)}_0(x,\tau) \right)^2
,
\\
\n{O(3)}
\frac{\pa \bar{\ph}^{(3)}_1(x,\tau)}{\pa \tau} &=&
- \left(- \pa^2 + \tilde{m}^2 \right) \bar{\ph}^{(3)}_1(x,\tau)
- \left(\xi - \frac16 \right) R \, \bar{\ph}^{(3)}_0(x,\tau)
- V^{(III)} \, \Bigl[  \bar{\ph}^{(1)}_0(x,\tau) \bar{\ph}^{(2)}_1(x,\tau)
\nonumber \\
&+&  \bar{\ph}^{(2)}_0(x,\tau) \bar{\ph}^{(1)}_1(x,\tau)
\Bigr] - \frac12  V^{(IV)} \, \left[\bar{\ph}^{(1)}_0(x,\tau) \right]^2 \bar{\ph}^{(1)}_1(x,\tau)
\nonumber \\
&-& \frac{1}{12} \, R_{\mu\nu} y^\mu y^\nu\, \left[  V^{(III)} \, \bar{\ph}^{(1)}_0(x,\tau)
\bar{\ph}^{(2)}_0(x,\tau)
+ \frac{1}{3}  V^{(IV)} \, \left(\bar{\ph}^{(1)}_0(x,\tau)\right)^3
\right],
\\
%O(4)
\n{C4}
\frac{\pa \bar{\ph}^{(4)}_1(x,\tau)}{\pa \tau} &=&
- \left(- \pa^2 + \tilde{m}^2 \right) \bar{\ph}^{(4)}_1(x,\tau)
- \left(\xi - \frac16 \right) R \, \bar{\ph}^{(4)}_0(x,\tau)
- V^{(III)} \, \Bigl[  \bar{\ph}^{(3)}_0(x,\tau) \bar{\ph}^{(1)}_1(x,\tau) \nonumber \\
&+&  \bar{\ph}^{(1)}_0(x,\tau) \bar{\ph}^{(3)}_1(x,\tau) + \bar{\ph}^{(2)}_0 \bar{\ph}^{(2)}_1(x,\tau)
\Bigr]
-   V^{(IV)} \, \Biggl[ \frac12 \left(\bar{\ph}^{(1)}_0(x,\tau) \right)^2 \bar{\ph}^{(2)}_1(x,\tau)
\nonumber \\
&+& \bar{\ph}^{(1)}_0(x,\tau) \bar{\ph}^{(2)}_0(x,\tau) \bar{\ph}^{(1)}_1(x,\tau)
\Biggr]
- \frac{1}{12} \, R_{\mu\nu} y^\mu y^\nu\,  \Bigg\{  V^{(III)} \,
\left[
\bar{\ph}^{(1)}_0(x,\tau) \bar{\ph}^{(3)}_0(x,\tau) + \frac12 \left(\bar{\ph}_0^{(2)}(x,\tau)\right)^2
\right] 
\nonumber 
\\
&+&  
V^{(IV)} \, \left(\bar{\ph}^{(1)}_0(x,\tau)\right)^2 \bar{\ph}_0^{(2)}(x,\tau)
\Bigg \} ,
\eeq
for curved spacetime up to first order in~$R$. We note that the last term in the expansion in
Eq.~(\ref{eboxg}) does not contribute to these equations because they vanish due to Lorentz
invariance~\cite{BP}. We note the presence of terms proportional to $R_{\mu\nu}$, which imply
that correlation functions beyond one-loop will not be be proportional to $(\xi-1/6)R$ only.

The corresponding curvature expansion of the one-loop
correlation function Eq.~\eq{ccn} is then given by
\beq
\label{ccn-e}
\bar{\De}^{\rm{1-loop}}(x,x'|\tau) =
\langle \bar{\ph}^{(1)}_0 (x,\tau)
\, \bar{\ph}^{(1)}_0 (x',\tau)\rangle_\eta
+ 2\, \langle
\bar{\ph}^{(1)}_0(x,\tau) \, \bar{\ph}^{(1)}_1(x',\tau)
\rangle_\eta
+ \cdots
\,,
\eeq
where we used the fact that $\langle
\bar{\ph}_0(x,\tau) \, \bar{\ph}_1(x',\tau)
\rangle_\eta =
\langle
\bar{\ph}_0(x',\tau) \, \bar{\ph}_1(x,\tau)
\rangle_\eta$
since  in the first order in curvature one can simply make
$x \leftrightarrow x'$ for the curvature expansions in
the normal coordinates, see, e.g., Eq.~\eq{eR}.
The first term in Eq.~(\ref{ccn-e}) corresponds to the
correlation function in flat spacetime which is already known,
while the second one is the first order in curvature correction
which is the new object to evaluate.

%%%%%%%%%%%%%% Solution of the equations %%%%
Equations~\eq{eq0} and \eq{C4} are solved analytically by using the Fourier
transforms of the scalar and noise fields,
\beq
\bar{\ph}^{(n)}_{0,1}(x,\tau)
= \int \frac{d^D k}{(2\pi)^D} \,
e^{ikx} \,\, \bar{\ph}^{(1)}_{0,1}(k,\tau)\,, \hspace{0.5cm}
\bar{\eta}(x,\tau)
=
\int \frac{d^D k}{(2\pi)^D} \,
e^{ikx} \,\, \bar{\eta} (k,\tau),
\label{FT-eta}
\eeq
with $\langle \bar{\eta}(k,\tau) \bar{\eta}(k',\tau')\rangle_\eta
=  2 \hbar  \, (2\pi)^D \delta^D(k+k') \delta(\tau - \tau')$, which
follows from Eq.~(\ref{mean-flat}). We exemplify the procedure for
Eqs.~\eq{eq0} and \eq{O(1)}, those required for the one-loop
correlation function:
\beq
\n{FL0}
\frac{\pa \bar{\ph}^{(1)}_0 (k,\tau)}{\pa \tau} &=&
- \left(k^2 + \tilde{m}^2\right) \bar{\ph}^{(1)}_0(k,\tau)
\,+\, \bar{\eta}(k,\tau)\,,
\\
\n{FL1}
\frac{\pa \bar{\ph}^{(1)}_1(k,\tau)}{\pa \tau} &=&
-\, \left(k^2 + \tilde{m}^2\right)\, \bar{\ph}^{(1)}_1(k,\tau)
- \left(\xi-\frac{1}{6}\right) R \, \bar{\ph}^{(1)}_0(k,\tau)\,.
\eeq
Using the initial conditions in Eqs.~\eq{ic0} and \eq{ic1}, one obtains:
\begin{eqnarray}
\label{S0}
\bar{\ph}^{(1)}_0(k,\tau) &=&
\int^{\tau}_{0} d \tau'
\, e^{- (k^2 + \tilde{m}^2)(\tau-\tau')}
\, \bar{\eta}(k,\tau'), \\
\label{Sleq1}
\bar{\ph}^{(1)}_1(k,\tau) &=& -\,\left(\xi-\frac{1}{6}\right) R
\int^{\tau}_{0} d\tau' \, e^{- (k^2 + \tilde{m}^2)(\tau-\tau')}
\, \bar{\ph}^{(1)}_0 (k,\tau')\,.
\end{eqnarray}
One can now evaluate the one-loop
correlation function: the flat space time part is given
by~\cite{Damgaard:1988nq}
\beq
\label{vm00}
\langle \bar{\ph}^{(1)}_0 (x,\tau) \,
\bar{\ph}^{(1)}_0 (x',\tau) \rangle_\eta =
\int \frac{d^D k}{(2\pi)^D} \, \frac{e^{iky}}{k^2 + \tilde{m}^2}
\left[1
\,-\, e^{-2(k^2 + \tilde{m}^2)\tau}
\right]\,,
\eeq
and the order-$R$ by
\begin{eqnarray}
\langle
\bar{\ph}^{(1)}_0 (k,\tau)
\, \bar{\ph}^{(1)}_1 (k',\tau)
\rangle_\eta &=&
- \left(\xi-\frac{1}{6}\right) R
\int \frac{d^D k}{(2\pi)^D} \, e^{iky} \left[
\frac{1- e^{-2(k^2 + \tilde{m}^2)\tau}}
{2(k^2 + \tilde{m}^2)^2}
-\frac{\tau \, e^{- 2(k^2 + \tilde{m}^2)\tau}}{k^2 + \tilde{m}^2}
\right]\,.
\label{DHc}
\end{eqnarray}
Therefore, $\bar{\De}^{1-loop}(x,x'|\tau)$ at first order in~$R$ is given by
\beq
\bar{\De}^{{\rm 1-loop}}(x,x'|\tau) &=&
\int \frac{d^D k}{(2\pi)^D} \, \,
e^{iky}\,
\left\{
\frac{1 - e^{-2(k^2 + \tilde{m}^2)\tau}}
{{k^2 +\tilde{m}^2}}
\right.
\nonumber \\
&-&
\left.
\Big(\xi-\frac{1}{6}\Big)R
\left[
\frac{1 - e^{-2(k^2 + \tilde{m}^2)\tau}}{(k^2 + \tilde{m}^2)^2}
- \frac{2\, \tau\, e^{- 2(k^2 + \tilde{m}^2)\tau}}{k^2 + \tilde{m}^2}
\right]
\right\} \,.
\label{Cft}
\eeq
In the equilibrium limit, one has
\beq
\n{GF}
\bar{G}(x,x') \,=\,
\lim_{\tau \to \infty} \bar{\De}^{{\rm 1-loop}}(x,x'|\tau)
\,=\, \int \frac{d^D k}{(2\pi)^D}
\, e^{iky}
\left[\frac{1}{k^2 + \tilde{m}^2}
- \Big( \xi-\frac{1}{6} \Big)
\frac{R}{(k^2 + \tilde{m}^2)^2}
\right],
\eeq
which reproduces the result of Bunch and Parker~\cite{BP}
for the momentum-space representation of the Feynman propagator.
This completes our first task, which is to derive a well known result
using stochastic quantization in curved space and normal coordinates.
Although this should not be taken by itself as the main result of the
present paper, it is nevertheless reassuring that one could obtain it
using a completely different method of stochastic quantization.

To conclude this section, we remark that the order in which the expansions
in  $\hbar$ and $R$ are made is not relevant. Moreover, for a
calculation with the full $\hbar$ dependence
retained, one performs only the expansion in~$R$.
Starting from Eq.~(\ref{le1}), redefining the scalar and noise fields
by $\ph = g^{-1/4}\, \bar{\ph}$ and $\eta = g^{-1/4}\, \bar{\eta}$
as previously, expanding $g=\det \big( g_{\mu\nu}\big)$ up to
the first order in~$R$ and writing
$\bar\ph = \bar\ph_0 + \bar\ph_1$, where $\bar\ph_0$
is ${\cal O}(R^0)$ and $\bar\ph_1$ is ${\cal O}(R^1)$,
one obtains the following equations for $\bar\ph_0$ and
$\bar\ph_1$:
\begin{eqnarray}
\frac{\pa \bar{\ph}_0(x,\tau)}{\pa \tau} &=&
\,-\, \left(-\pa^2 + m^2\right) \bar{\ph}_0(x,\tau)
- \, \lambda \, {\bar\ph}^3_0(x,\tau) + \bar{\eta}(x,\tau) ,
\label{eq-phi0} \\
\frac{\pa \bar{\ph}_1(x,\tau)}{\pa \tau} &=&
\,-\, \left(-\pa^2 + m^2\right) \bar{\ph}_1(x,\tau)
- \Big(\xi - \frac{1}{6} \Big)\,R \, \bar{\ph}_0(x,\tau)
\nonumber
\\
&&
- \, 3 \lambda \, {\bar\ph}^2_0(x,\tau){\bar\ph}_1(x,\tau)
- \frac{1}{6}\, R_{\mu\nu} y^\mu y^\nu \,  \lambda
\, {\bar\ph}^3_0(x,\tau).
\label{eq-phi1}
\end{eqnarray}
Clearly, these equations contain the full $\hbar$ dependence. When $\bar\ph_0$ and $\bar\ph_1$ are
expanded as in Eq.~(\ref{en}), one recovers the loop expansion
discussed above. In section~\ref{sect4} we present comparisons of analytical and numerical
solutions.

%%%%%%%%%%%%%%%%%%%%%%%%%%%%
\subsection{Effective potential}
\label{sect3}

After deriving Eq.~(\ref{GF}), one can use the result for calculating
the effective potential. The effective potential $V_{eff} (\phi)$ is
defined as the zeroth-order term in the derivative expansion of the
effective action of a background scalar field $\phi(x)$:
\beq
V_{eff} (\phi) =  \frac12\, m^2 \phi^2
- \frac12\, \xi R \phi^2 + V(\phi)
+ \bar{V}^{(1)} (\phi)
+ \cdots\,,
\eeq
where $\bar{V}^{(1)} (\phi)$ is the one-loop correction to
the effective potential. The curvature expansion of
$\bar{V}^{(1)} (\phi)$ reads
$\bar{V}^{(1)}  = \bar{V}^{(1)}_0  \,+\, \bar{V}^{(1)}_1 \,+\, \cdots $,
where $\bar{V}^{(1)}_0$ is the well-known flat spacetime effective
potential, which has been derived many times and in different ways
starting from the work of Coleman and Weinberg~\cite{CoWe}.
The order$-R$ correction $\bar{V}^{(1)}_1(\phi)$, as shown
{\it e.g.} in Ref.~\cite{Sobreira:2011ep}, can be written in terms
of $\bar{G}_{0}(k)$ and $\bar{G}_{1}(k)$, the flat spacetime
and order$-R$ terms in the integrand of Eq.~\eq{GF}.

The result for the effective potential is
\beq
\bar{V}_0^{(1)}(\phi)
&=&
\frac12 \int \frac{d^D k}{(2\pi)^D} \,
\ln \left( \frac{k^2+m^2+V''}{k^2+m^2}  \right)
\,\,=\,\,
\frac{\Om^D}{2^{D+1} \pi^{D/2}}
\Bigg\{
\frac{1}{\Ga \big(\frac{D}{2}+1\big)}
\ln \left(\frac{\Om^2 + m^2+V''}{\Om^2+m^2}\right)
\n{V0-sol}
\\
&-&
\frac{\Om^2}{\Ga\big(\frac{D}{2}+2\big)}
\Bigg[
\frac{1}{m^2+V''} \,
{}_2 F_1 \left(1,\tfrac{D+2}{2},\tfrac{D+4}{2},
-\frac{\Om^2}{m^2+V''}\right)
-
\frac{1}{m^2}
\,{}_2 F_1 \left(1,\tfrac{D+2}{2},\tfrac{D+4}{2},
-\frac{\Omega ^2}{m^2}\right)\Bigg]\Bigg\},
\nonumber
\eeq
for the flat space part\footnote{The last term in this expression
is a vacuum contribution, for it does not depend on $\phi$.} and
\beq
\n{GHR1}
\bar{V}^{(1)}_1 (\phi) &=& - \frac{1}{2} \int \, \frac{d^D k}{(2\pi)^D} \,
\bar{G}^{-1}_{0}(k) \, \bar{G}_{1}(k)
= \frac{1}{2} \,  \left( \xi-\frac{1}{6} \right)R
\int \, \frac{d^D k}{(2\pi)^D}
\frac{1}{k^2 + m^2 + V''}
\nonumber
\\
&=&
\left( \xi-\frac{1}{6} \right)R\,\, \frac{\Om^D}{2^D \pi^{D/2} \, D \,
\Ga(\tfrac{D}{2})} \frac{1}{m^2+V''}\,
{}_2F_1\left( 1,\tfrac{D}{2};\tfrac{D+2}{2};
-\frac{\Omega^2}{m^2+V''} \right),
\end{eqnarray}
for the first order terms.
For $D=4$ both expressions precisely give the result obtained
in Ref.~\cite{Sobreira:2011ep}, where the renormalization and
related aspects were discussed in detail and hence will not be
considered here.

%%%%%%%%%%%%%%%%%%%%%%%%%%%%%%%%
%%%%%%%%%%%%%%%%%%%%%%%%%%%%%%%%
\section{Numerical simulations}
\label{sect4}

The main advantage of the approach based on the Riemann normal
coordinates is that the practical calculation is performed in flat
space. In particular, this means we can use the well-known methods of
lattice-regularized Langevin numerical simulations. It should be clear,
however, that the results have validity for small $R$ and $y^\mu$.

For this initial investigation, we perform numerical simulations for
the simpler case of $D=3$; therefore, the mass dimensions of the
relevant quantities are as follows:
\ $[\bar\varphi] = 1/2$, \ $[\lambda] = 1$,
\ $[\tau] = -2$, \ $[\bar\eta] = 5/2$, and \
$[R] = 2$. We solve the Langevin equations on a $N^3$ lattice, with
lattice spacing $a$. The Langevin-time discretization is denoted
$\Delta\tau$. It is convenient to rescale all dimensionful quantities
by $a$, namely
\beq
\bar\varphi = \hat{\phi}\,a^{-1/2},
\quad
x =\hat x\,a,
\quad
\tilde{m} = \hat{m}\,a^{-1},
\quad
\lambda = \hat\lambda\,a^{-1},
\quad
\bar\eta = \hat\eta \,a^{-5/2}\, \epsilon^{1/2},
\quad
\tau = \hat\tau\,a^2,
\quad
\Delta\tau = \epsilon \, a^2.
\eeq
We exemplify the numerical procedure for Eqs.~(\ref{eq0}) and (\ref{O(1)}). In terms of these
rescaled quantities, the corresponding discretized equations are given (in It\^o calculus~\cite{Gardiner}) by
\begin{eqnarray}
\hat\phi^{(1)}_0(\hat x,\hat\tau + \epsilon)
&=&
\hat\phi^{(1)}_0(\hat x,\hat\tau)
+ \epsilon\, \left({\hat\square} - {\hat m}^2\right)
\hat\phi^{(1)}_0(\hat x,\hat \tau)
+ \sqrt{\epsilon} \, \hat\eta(\hat x,\hat\tau),
\label{eq0-disc}
\\
\hat\phi^{(1)}_1(\hat x,\hat\tau + \epsilon)
&=&
\hat\phi^{(1)}_1(\hat x,\hat\tau)
+ \epsilon\, \left({\hat\square} - {\hat m}^2\right)
\hat\phi^{(1)}_1(\hat x,\hat \tau)
- \left(\xi - \frac{1}{6}\right) \hat R
\, \hat\phi^{(1)}_0(\hat x,\hat \tau) ,
\label{eq1-disc}
\end{eqnarray}
with \ $\hat\square \, \hat\phi = \hat\phi(\hat x + \hat e) - 2 \hat\phi(\hat x)
+ \hat\phi(\hat x - \hat e)$, where $\hat e$ is a unit vector, and the noise
correlation is $
\langle\hat\eta(\hat x,\hat\tau) \, \hat\eta(\hat x',\hat\tau')\rangle_\eta=
2 \, \delta_{\hat x,{\hat x}'} \, \delta_{\hat \tau,{\hat\tau}'}$.
The other equations for $\hat\phi^{(2)}_0$, $\hat\phi^{(2)}_1$,
$\dots$, as well as the nonperturbative equations in Eqs.~(\ref{eq-phi0})
and (\ref{eq-phi1}), are discretized in the same manner.

We solve all Langevin equations with $\epsilon = 10^{-4}$ on a $N^3 = 16^3$ lattice.
The values of the lattice mass and coupling constant are chosen
$\hat m = 1 = \hat\lambda$. The scalar curvature $R$ is supposed to be small
for the expansion in $R$ to make sense; therefore, we choose $\hat R = 0.1$.
The classical background field in flat spacetime is the constant field at the
minimum of the classical potential, $\hat\phi_{\rm cl} = \sqrt{-\hat m^2/{\hat\lambda}} = 1$.
The corresponding bare ``Higgs mass'' is \ ${\hat m}^2_H = - {\hat m}^2
+ 3 {\hat \lambda} \hat\phi^2_{\rm cl} = 2 {\hat m}^2 = 2$.

Let us first consider the one-loop correlation function. We concentrate on the
$R-$dependent part of the correlation function, the second term in Eq.~(\ref{ccn-e}).
In terms of lattice variables, it is given by
\begin{eqnarray}
\hat\Delta^{{\rm 1-loop}}_R(\hat x,\hat x|\tau)
&=&
2 \,\langle \hat\phi^{(1)}_0(\hat x,\hat\tau)\,
\hat\phi^{(1)}_1(\hat x,\hat\tau) \rangle_{\hat\eta} \nonumber \\
&=&
- \,\left( \xi - \frac{1}{6}\right) \frac{{\hat R}}{16 N^3}
\sum^{N-1}_{n_1, n_2, n_3=0}
\left\{\frac{1-e^{-2\hat\tau({\hat k}^2 + {\hat m}^2)}}
{\left[d(\hat m,n_1,n_2,n_3)\right]^2}
- \frac{2\hat\tau e^{-2\hat\tau({\hat k}^2
+ {\hat m}^2)}}{d(\hat m,n_1,n_2,n_3)} \,
\right\},
\label{hatDelta-1loop}
\end{eqnarray}
where $\,d(\hat m, n_1,n_2,n_3) = ({\hat m}/2)^2 + \sum^3_{i=1}
\sin^2 (n_i \pi/N)$. This expression corresponds to
the limit $\tau \to \infty$, the equilibrium from the viewpoint of
stochastic quantization. The one-loop expression has an overall factor of
$\xi-1/6$. This factor is a common feature in the one-loop
results for the nonminimal term. We have performed a lengthy two-loop
calculation to obtain the two-loop generalization\footnote{We expect
to report it soon in a separate more extensive publication~\cite{us-expl}.} of
Eq.~(\ref{hatDelta-1loop}) and found that it is not proportional
to $\xi-1/6$ only, because of the $R_{\mu\nu}$ term in Eqs.~(\ref{O(2)})-(\ref{C4}),
regardless of the fact that in the three-dimensional space there is no conformal
symmetry at $\xi=1/6$, no one-loop divergences and hence there is
no corresponding anomaly.

\begin{figure*}[t]
\begin{center}
\includegraphics[height=6.0cm,angle=0]{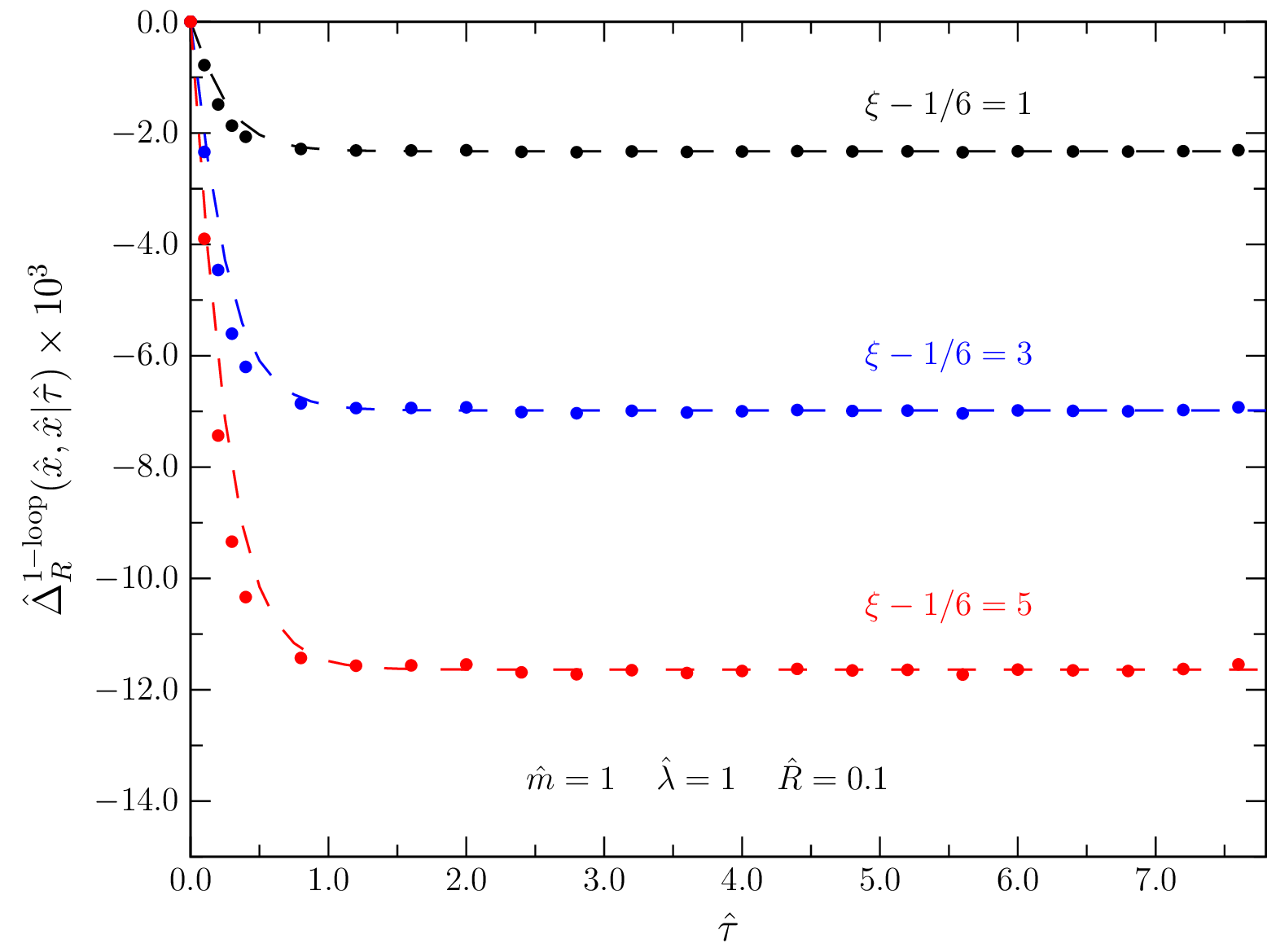}\hspace{0.4cm}
\caption{$R-$dependent part of the correlation function, second term
in Eq.~(\ref{ccn-e}), as a function of $\hat\tau$. The dotted points
are the results of the numerical simulations and the dashed lined lines correspond to the
analytic result, Eq.~(\ref{hatDelta-1loop}).
}
\label{fig1}
\end{center}
\end{figure*}

Figure~\ref{fig1} shows the one-loop results of the
numerical simulations (with stochastic averages taken over 100
noise realizations) for the $R-$dependent part of the correlation
function for different values of
$(\xi - 1/6)$. The corresponding analytical results are also shown by the dashed lines.
The figure reveals that the lattice simulations reproduce very well the
analytical results. We note that changing the values of the
parameters does not spoil this good agreement.

\begin{figure*}[t]
\begin{center}
\includegraphics[height=6.0cm,angle=0]{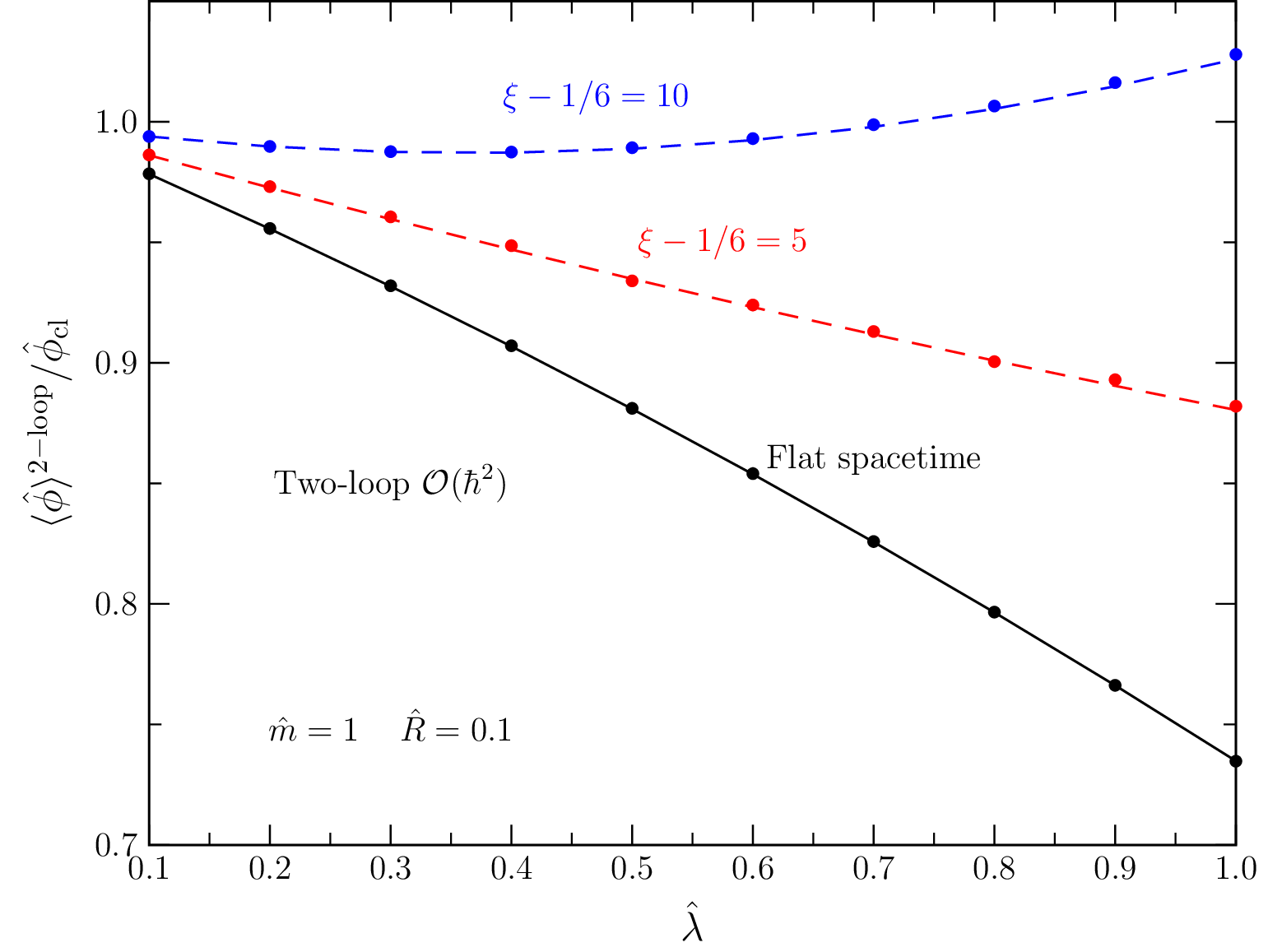}\hspace{0.4cm}
\includegraphics[height=6.0cm,angle=0]{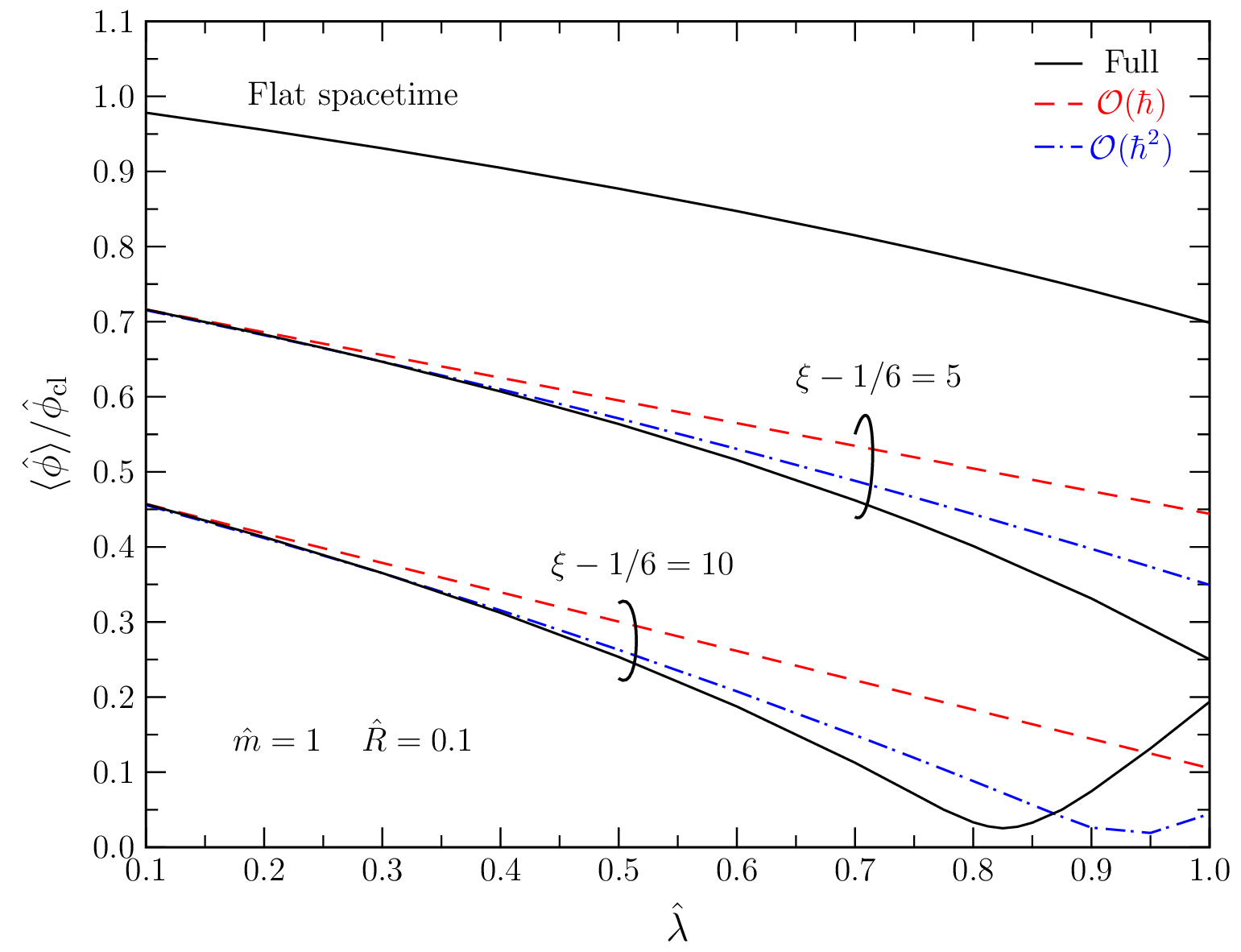}
\caption{Vacuum expectation value of the field, Eq.~(\ref{av_hat-phi}), as function
of the coupling, and normalized to $\hat\phi_{\rm cl} = \sqrt{-{\hat m}^2/\hat\lambda}$.
Left panel: Analytical and numerical results up to two loops, Eq.~(\ref{phi-2loop}),
for the flat and curved spacetimes, with a  flat and homogeneous classical background.
The dotted points are the results of the
numerical simulations. Right panel: All solid (black) curves represent the solutions of the full
Langevin equations, Eqs.~(\ref{eq-phi0}): the uppermost curve is for flat spacetime ($\hat R = 0$)
and the lower ones for curved spacetime ($\hat R = 0.1$), the classical background includes
curvature effects.}
\label{fig2}
\end{center}
\end{figure*}

Next, we discuss results for the vacuum expectation value
$\langle\hat \phi\rangle$ of the field (recall we are considering the case
of spontaneous symmetry breaking):
\beq
\langle \hat\phi \rangle
= \lim_{\tau \rightarrow \infty} \frac{1}{N^3} \sum^{N-1}_{n_1, n_2, n_3=0}
\langle \hat\phi(\hat x,\hat\tau) \rangle_\eta.
\label{av_hat-phi}
\eeq
We choose the positive values of $\langle\hat \phi\rangle$, as the action
has the $\ph \rightarrow - \ph$ symmetry.
Initially we compare analytical and numerical calculations. The expressions for
the analytic results, particularly those at two loops, are too lengthly to be shown here and
will be presented elsewhere~\cite{us-expl}. In the left panel of Fig.~\ref{fig2}
we show results for $\langle \hat\phi \rangle$ as a function of the coupling $\hat \lambda$
calculated up to two loops:
\beq
\langle \hat\phi \rangle^{\rm{2-loop}} = \left( \hat \phi_{\rm cl} + \langle \hat \phi^{(2)}_0 \rangle
+ \langle \hat \phi^{(2)}_1 \rangle + \langle \hat \phi^{(4)}_0 \rangle
+ \langle \hat \phi^{(4)}_1 \rangle
\right)/\hat \phi_{\rm cl}.
\label{phi-2loop}
\eeq
For this comparison, the analytical (and therefore the numerical as well)
results are for a homogeneous and flat classical background, i.e.
$\hat\phi_{\rm cl} = \sqrt{-{\hat m}^2/\hat\lambda}$. Terms proportional to $R_{\mu\nu}$
have been neglected as well. The agreement between analytical and numerical (shown as
dots in the figure) results is very good. The agreement is now spoiled for other values
of $\hat m$ and $\hat R$ and also $\xi$.

Finally, we discuss results for $\langle\hat \phi\rangle$ with the full dependence
on $\hbar$ retained, i.e. results obtained from the solutions of Eq.~(\ref{eq-phi0})
and~(\ref{eq-phi1}). We also compare with the corresponding one- and two-loop results.
Now we include the term $R_{\mu\nu} y^\mu y^\nu$ and consider curvature effects on
the classical background. Since $y^\mu$ must be small, we take it as being the lattice
spacing{\textemdash}smaller values of $y^\mu$ would make the last term
in Eq.~(\ref{eq-phi1}) even less important. For $R_{\mu\nu}$ we invoke rotational
symmetry and take $R_{\mu\nu} = R_{11} = 1/3 \, R$ for all $\mu$ and~$\nu$. The results are
shown on the right panel of Fig.~\ref{fig2}. The full$-\hbar$ solutions are the solid (black)
curves, and one- and two-loop solutions are the (red) dashed and (blue) dot-dashed curves,
respectively. Not surprisingly, the figure reveals that as $\hat\lambda$ increases, the ${\cal O}(\hbar)$
results deviate from the nonperturbative ones considerably. But it also
reveals that there is improvement when going to ${\cal O}(\hbar^2)$, at
least for the parameters chosen. There is an interesting interplay between the roles
played by $\lambda$ and $\xi$, as indicated by the presence of the dip at $\hat\lambda
\sim 0.8$ in the solutions for $(\xi - 1/6) = 10$. This is due to the fact that the equilibrium
values of $\ph_0$ and $\ph_1$ have opposite signs. This is clear for the classical
solutions of Eqs.~(\ref{eq-phi0}) and (\ref{eq-phi1}): while $(\bar\ph_0)_{\rm cl} =
\pm \sqrt{-m^2/\lambda}$, the order$-R$ solution is
$(\bar\ph_1)_{\rm cl} = - [(\xi - 1/6)R/(m^2+ 3\lambda\ph_0)]\, (\bar\ph_0)_{\rm cl}$,
when neglecting the  smaller term proportional do $R_{\mu\nu}$ in Eq.~(\ref{eq-phi1}).
For larger values of $\hat\lambda$, quantum fluctuations make the $\hat\phi_0$
contribution dominate over $\hat\phi_1$, as can be seen by the rising of the
curve towards the flat-spacetime solution. We note that the dip is also present in the
curves for $(\xi - 1/6)R$, but for a value of $\hat\lambda$ beyond the range shown
in the figure.

%%%%%%%%%%%%%%%%%%%%%%%%%%%%%%%%
%%%%%%%%%%%%%%%%%%%%%%%%%%%%%%%%
\section{Conclusions and Perspectives}
\label{sect5}

The formalism of stochastic quantization enables one to go beyond
the scope of the usual perturbation theory, in particular when using
numerical lattice methods for solving the associated Langevin equation.
We presented a construction of this equation for the self-interacting
scalar field in an arbitrary curved background. The solution of the
Langevin equation can be carried on either analytically or numerically,
by means of the local momentum representation.

In the analytical part of the work we used the Langevin equation to
reproduce the known result for the effective potential of
self-interacting scalar field in curved space in the dimension $D=4$,
confirming the that the equivalence of stochastic quantization
and path integral also holds in curved space. Furthermore, we derive
the effective potential at one loop order in an arbitrary dimension  $D$
in curved space. To the best of our knowledge this result is original. We
have also made of analytical and numerical two-loop calculations.

The main point of the paper is the complementarity of the analytic
and numerical approaches, especially the possibility of using numerical
simulations based on lattice methods in curved spacetime. This
is facilitated by the local momentum representation, which allows us
to obtain curved-space results by making calculations in flat space,
associated with a given point $P$.

From the exercise described in Sec.~\ref{sect4} it is clear that the
numerical simulations can be carried out to an arbitrary order in
$\hbar$, a feature that is hardly possible in analytical calculations
due to the appearance of multiloop integrals at higher orders that
become more and more complicated to solve. The numerical solution is
facilitated because the perturbative Langevin equations for the different
orders in $\hbar$ have the same structure, they are linear and can
be solved iteratively. Furthermore, as we have shown explicitly, the use of
the loop expansion is actually not necessary at all, as
numerical simulations can be performed using an expansion in~$R$
only and not in $\hbar$, i.e. no expansion
in the noise field. Again, to our knowledge such results
have been obtained for the first time in the present paper.

Given the limited scope of this first publication regarding numerical
simulations, we do not pursue such an analyses further here. It is
planned to make a more complete analysis in a future publication
for the realistic case of $D=4$ for which, in particular, the continuum
limit and renormalization will be studied. This will allow detailed comparisons
with and extension to higher-orders of previous two-loop calculations,
e.g. those of Refs.~\cite{Hu:1984js,Tsamis:1997za}.

The progress in nonperturbative methods in curved spacetime
would pave the way for future work. Perhaps the most
relevant step would be the development of nonperturbative
methods of evaluating the effective action, which is a generalization
of effective potential for the non-constant background field.
Another possible development of the approach which we
presented above is related to the gauge-independent quantization
of the theories with unbroken, softly broken and even strongly
broken gauge symmetries. We expect to deal with those issues in
future works.

%%%%%%%%%%%%%%%%%%%%%%%%%%%%%
%%%%%%%%%%%%%%%%%%%%%%%%%%%%%
\vskip 0.2cm \noindent
{\bf Acknowledgments} \\
This work was partially supported by Conselho Nacional de
Desenvolvimento Cient\'{i}fico e Tecnol\'{o}gico - CNPq,
305894/2009-9 (G.K.), 464898/2014-5 (G.K., INCT F\'{\i}sica
Nuclear e Apli\-ca\-\c{c}\~oes), 303893/2014-1 (I.Sh.),
Funda\c{c}\~{a}o de Amparo \`{a} Pesquisa do Estado de S\~{a}o
Paulo - FAPESP, 2013/01907-0 (G.K.), and Funda\c{c}\~{a}o de Amparo
\`a Pesquisa de Minas Gerais - FAPEMIG, APQ-01205-16 (I.Sh.).
E.A.R. is grateful to Coordena\c{c}\~ao de Aperfei\c{c}oamento de Pessoal
de N\'{\i}vel Superior - CAPES  for supporting his Ph.D. project.
The work of  was supported by the PNPD program from CAPES.
%%%%%%%%%%%%%%%%%%%%%%%%%%%%%%%%%%%%

%%%%%%%%%%%%%%%%%%%%%%%%%%%
\end{document}